\documentstyle[graphicx,aps,multicol]{revtex}
\begin{document}
\draft
\title{Microcanonical entropies and radiative strength functions of $^{50,51}$V}
\author{A.C.~Larsen\footnote{Electronic address: a.c.larsen@fys.uio.no}, 
R.~Chankova, M.~Guttormsen, F.~Ingebretsen, S.~Messelt, J.~Rekstad, S.~Siem, N.U.H.~Syed, and S.W.~{\O}deg{\aa}rd}
\address{Department of Physics, University of Oslo, P.O.Box 1048 Blindern, N-0316 Oslo, Norway}

\author{ T.~L\"{o}nnroth}
\address{Department of Physics, \AA bo Akademi University, 
FIN-20500 \AA bo, Finland}

\author{ A.~Schiller}
\address{National Superconducting Cyclotron Laboratory, Michigan State University, East Lansing, MI 48824, USA}

\author{A.~Voinov}
\address{Department of Physics and Astronomy, Ohio University, Athens, Ohio 45701, USA}
\maketitle

\begin{abstract}
The level densities and radiative strength functions (RSFs) of $^{50,51}$V have been extracted using the ($^3$He,$\alpha \gamma$) and ($^3$He,$^3$He$^{\prime} \gamma$) reactions, respectively. From the level densities, microcanonical entropies are deduced. The high $\gamma$-energy part of the measured RSF fits well with the tail of the giant electric dipole resonance. A significant enhancement over the predicted strength in the region of $E_{\gamma} \lesssim 3$ MeV is seen, which at present has no theoretical explanation.
\end{abstract} 


\begin{multicols}{2}

\section{Introduction}

The structure of the vanadium isotopes is based on simple shell-model configurations at low excitation energies. The valence protons and neutrons are occupying the single-particle $\pi f_{7/2}$ and $\nu f_{7/2}$ orbitals, respectively. These shells are isolated from other orbitals by the {\it N}, {\it Z} = 20 and 28 shell gaps, making the vanadium isotopes interesting objects for studying various nuclear shell effects. In particular, it is well known that the number of available singe-particle levels is significantly reduced for nuclei at closed shells. 

The density of states or, equivalently, the entropy in these systems depends on the number of broken Cooper pairs and single-particle orbitals made available by crossing the shell gaps. The $^{50,51}$V nuclei are of special interest because the neutrons are strongly blocked in the process of creating entropy; $^{50}$V and $^{51}$V have seven and eight neutrons in the $\nu f_{7/2}$ orbital, respectively. Thus, the configuration space of the three protons in the $\pi f_{7/2}$ shell is of great importance. 

These particular shell-model configurations are also expected to govern the $\gamma$-decay routes. Specifically, as within every major shell, the presence of only one parity for single-particle orbitals in the low-spin domain means that transitions of $E1$ type will be suppressed. The low mass of the investigated nuclei causes that the centroid of the giant electric dipole resonance (GEDR) is relatively high while the integrated strength according to the Thomas-Reiche-Kuhn sum rule is low, both observations working together to produce a relatively weak low-energy tail when compared to heavier nuclei. Hence, possible non-statistical effects in the radiative strength function (RSF) might stand out more in the present investigation. 

The Oslo Cyclotron group has developed a method to extract first-generation (primary) $\gamma$-ray spectra at various initial excitation energies. From such a set of primary spectra, the nuclear level density and the RSF can be extracted simultaneously \cite{hend1,schi0}. These two quantities reveal essential information on nuclear structure such as pair correlations and thermal and electromagnetic properties. In the last five years, the Oslo group has demonstrated several fruitful applications of the method \cite{melb0,schi2,gutt3,voin1,siem1}.

In Sect.~II an outline of the experimental procedure is given. The level densities and microcanonical entropies are discussed in Sect.~III, and in Sect.~IV the RSFs are presented. Finally, concluding remarks are given in Sect.~V.

\section{Experimental method}

The experiment was carried out at the Oslo Cyclotron Laboratory (OCL) using a beam of 30-MeV $^3$He ions. The self-supporting natural V target had a purity of $99.8$\% and a thickness of 2.3 mg/cm$^2$. Particle-$\gamma$ coincidences for $^{50,51}$V were measured with the CACTUS multi-detector array \cite{Cactus}. The charged ejectiles were detected using eight Si particle telescopes placed at an angle of 45$^{\circ}$ relative to the beam direction. Each telescope consists of a front $\Delta E$ detector and a back $E$ detector with thicknesses of 140 and 1500 $\mu$m, respectively. An array of 28 collimated NaI $\gamma$-ray detectors with a total efficiency of $\sim$15\% surrounded the target and the particle detectors. The reactions of interest were the pick-up reaction $^{51}$V($^3$He,$\alpha \gamma$)$^{50}$V, and the inelastic scattering $^{51}$V($^3$He,$^3$He$^{\prime}\gamma$)$^{51}$V. The typical spin range is expected to be $I\sim 2-4 \hbar$. The experiment ran for about one week, with beam currents of $\sim 1$ nA.

The experimental extraction procedure and the assumptions made are described in Refs.~\cite{hend1,schi0}. 
The data analysis is based on three main steps:
\begin{itemize}
\item[(1)] preparing the particle-$\gamma$ coincidence matrix
\item[(2)] unfolding the $\gamma$-ray spectra
\item[(3)] constructing the first-generation matrix 
\end{itemize}

In the first step, for each particle-energy bin, total spectra of the $\gamma$-ray cascades are obtained from the coincidence measurement. The particle energy measured in the telescopes is transformed to excitation energy of the residual nucleus, using the reaction kinematics. Then each row of the coincidence matrix corresponds to a certain excitation energy $E$, while each column corresponds to a certain $\gamma$ energy $E_\gamma$. 

\begin{figure}
\includegraphics[totalheight=21cm,angle=0,bb=0 80 350 730]{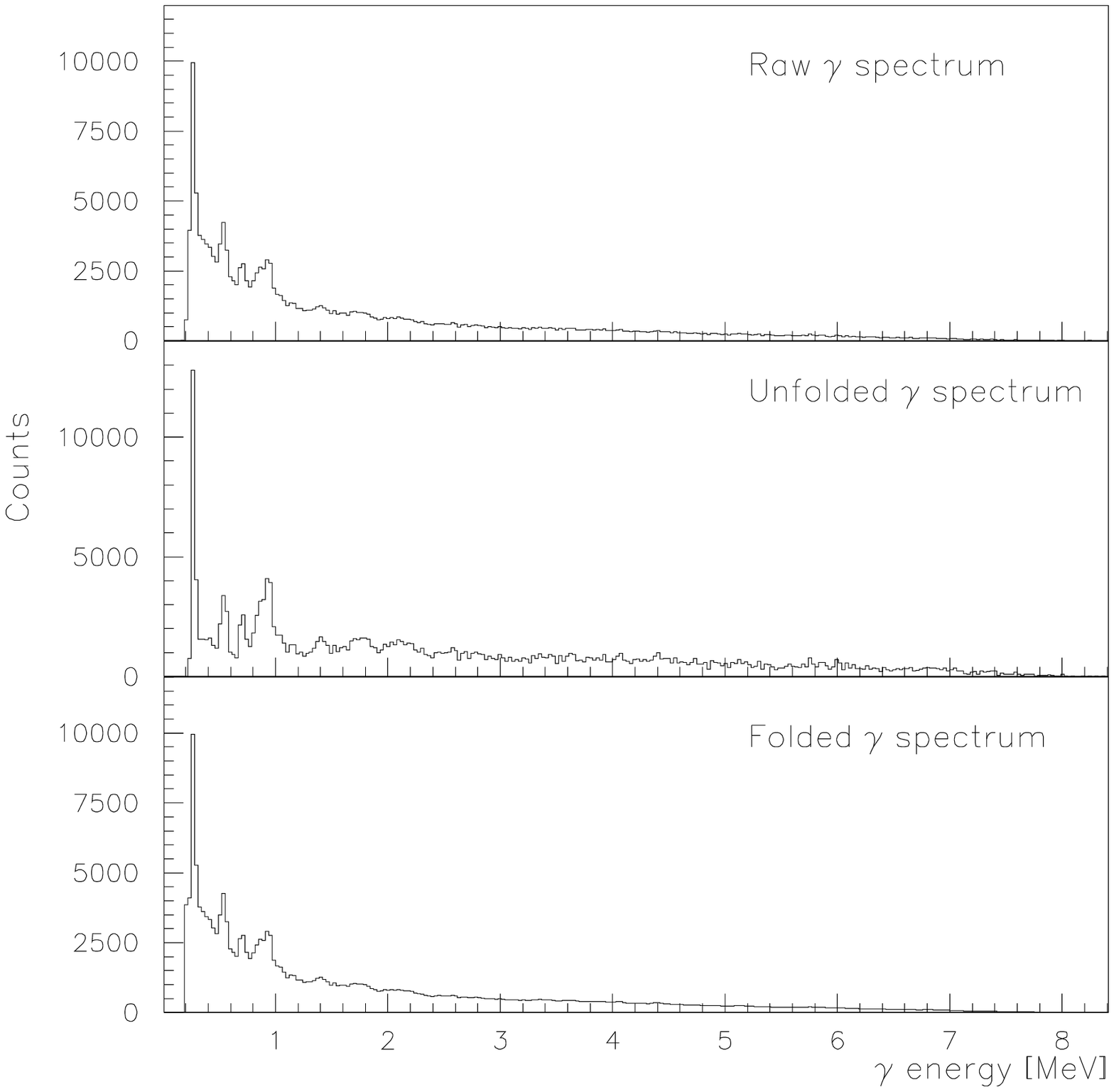}
\caption{$\gamma$ spectra of $^{50}$V for excitation energy $E = 6 - 8$ MeV.}
\label{fig:unfolding}
\end{figure}

In the next step, the $\gamma$-ray spectra are unfolded using the known response functions of the CACTUS array~\cite{gutt6}. The Compton-subtraction method described in Ref.~\cite{gutt6} preserves the fluctuations in the original spectra without introducing further, spurious fluctuations. A typical raw $\gamma$ spectrum is shown in the top panel of Fig.~\ref{fig:unfolding}, taken from the $^{50}$V coincidence matrix gating on the excitation energies between $E = 6 - 8$ MeV. The middle panel shows the unfolded spectrum, and in the bottom panel this spectrum has been folded with the response functions. The top and bottom panels are in excellent agreement, indicating that the unfolding method works very well.  

\begin{figure}
\includegraphics[totalheight=21cm,angle=0,bb=0 80 350 730]{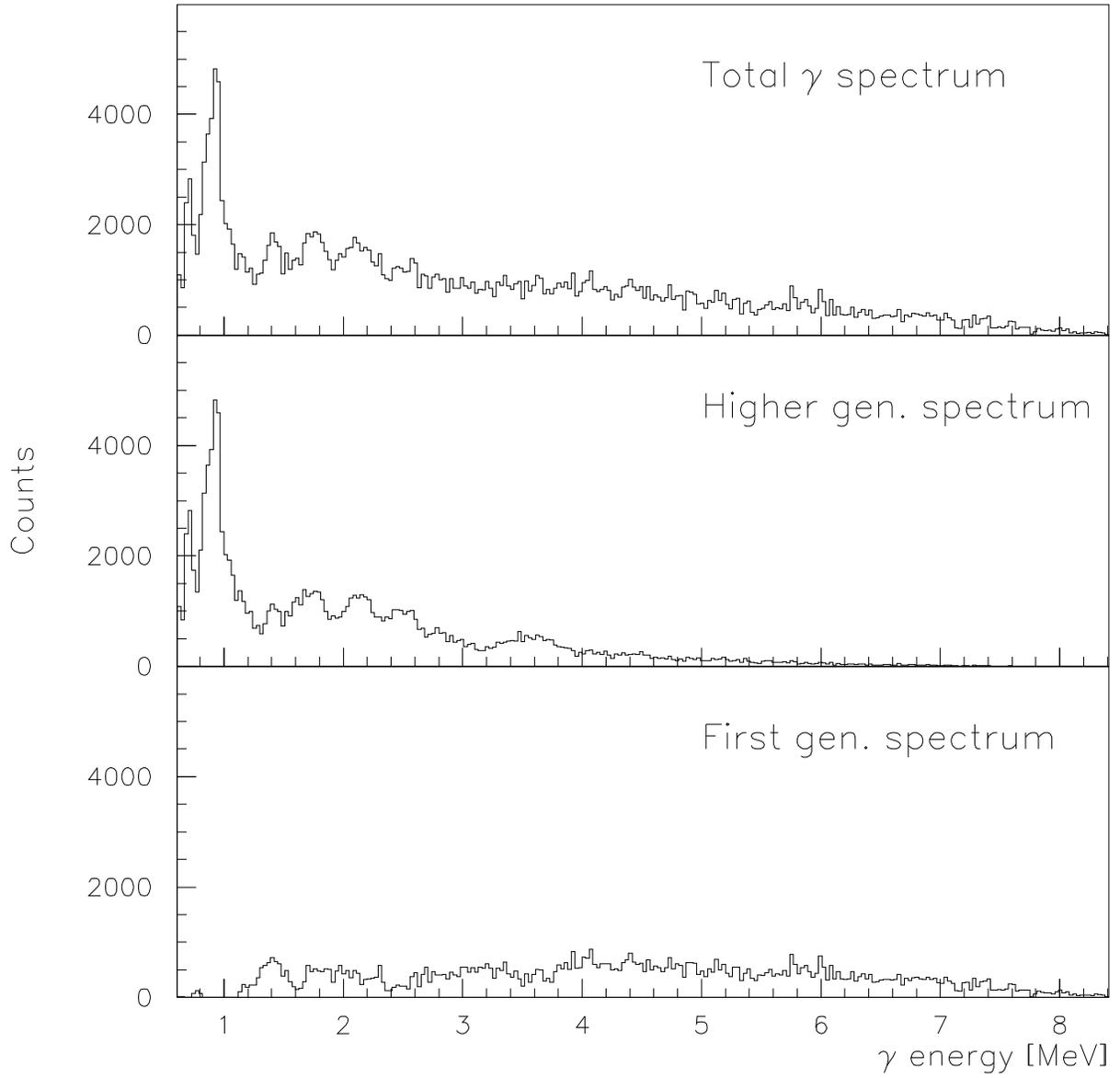}
\caption{Unfolded $\gamma$ spectra of $^{50}$V for excitation energy $E = 6 - 8$ MeV.}
\label{fig:gamma}
\end{figure}

The third step is to extract the $\gamma$-ray spectra containing only the first $\gamma$ rays in a cascade. These spectra are obtained for each excitation-energy bin through a subtraction procedure as described in Ref.~\cite{gutt0}. The main assumption of this method is that the $\gamma$-decay spectrum from any excitation-energy bin is independent of the method of formation, either directly by the nuclear reaction or populated by $\gamma$ decay from higher-lying states following the initial reaction. This assumption is automatically fulfilled when the same states are equally populated by the two processes, since $\gamma$ branching ratios are properties of the levels themselves. Even if different states are populated, the assumption is still valid for statistical $\gamma$ decay, which only depends on the $\gamma$-ray energy and the number of accessible final states. In Fig.~\ref{fig:gamma}, the total, unfolded $\gamma$ spectrum, the second and higher generations $\gamma$ spectrum and the first-generation spectrum of $^{50}$V with excitation-energy gates $E = 6 - 8$ MeV are shown. The first-generation spectrum is obtained by subtracting the higher-generation $\gamma$ rays from the total $\gamma$ spectrum. 

When the first-generation matrix is properly normalized~\cite{schi0}, the entries of it are the probabilities $P(E, E_{\gamma})$ that a $\gamma$-ray of energy $E_{\gamma}$ is emitted from an excitation energy $E$. The probability of $\gamma$ decay is proportional to the product of the level density $\rho (E-E_{\gamma})$ at the final energy $E_{f} = E-E_{\gamma}$ and the $\gamma$-ray transmission coefficient ${\mathcal T}(E_{\gamma})$:

\begin{equation}
P(E, E_{\gamma}) \propto  \rho (E -E_{\gamma}) {\mathcal{T}}  (E_{\gamma}).
\label{eq:axel}
\end{equation}
This factorization is the generalized form of the Brink-Axel hypothesis\cite{brink,axel}, which states that any excitation modes built on excited states have the same properties as those built on the ground state. This means that the $\gamma$-ray transmission coefficient is independent of excitation energy and thus of the nuclear temperature of the excited states. There is evidence that the width of the giant dipole resonance varies with the nuclear temperature of the state on which it is built~\cite{kad,Ger}. However, the temperature corresponding to the excitation-energy range covered in this work is rather low and changes slowly with excitation energy ($T \sim \sqrt{E_{f}}$ ); thus we assume a constant temperature and that the $\gamma$-ray transmission coefficient does not depend on the excitation energy in the energy interval under consideration. 

\begin{figure}
\includegraphics[totalheight=21cm,angle=0,bb=0 80 350 730]{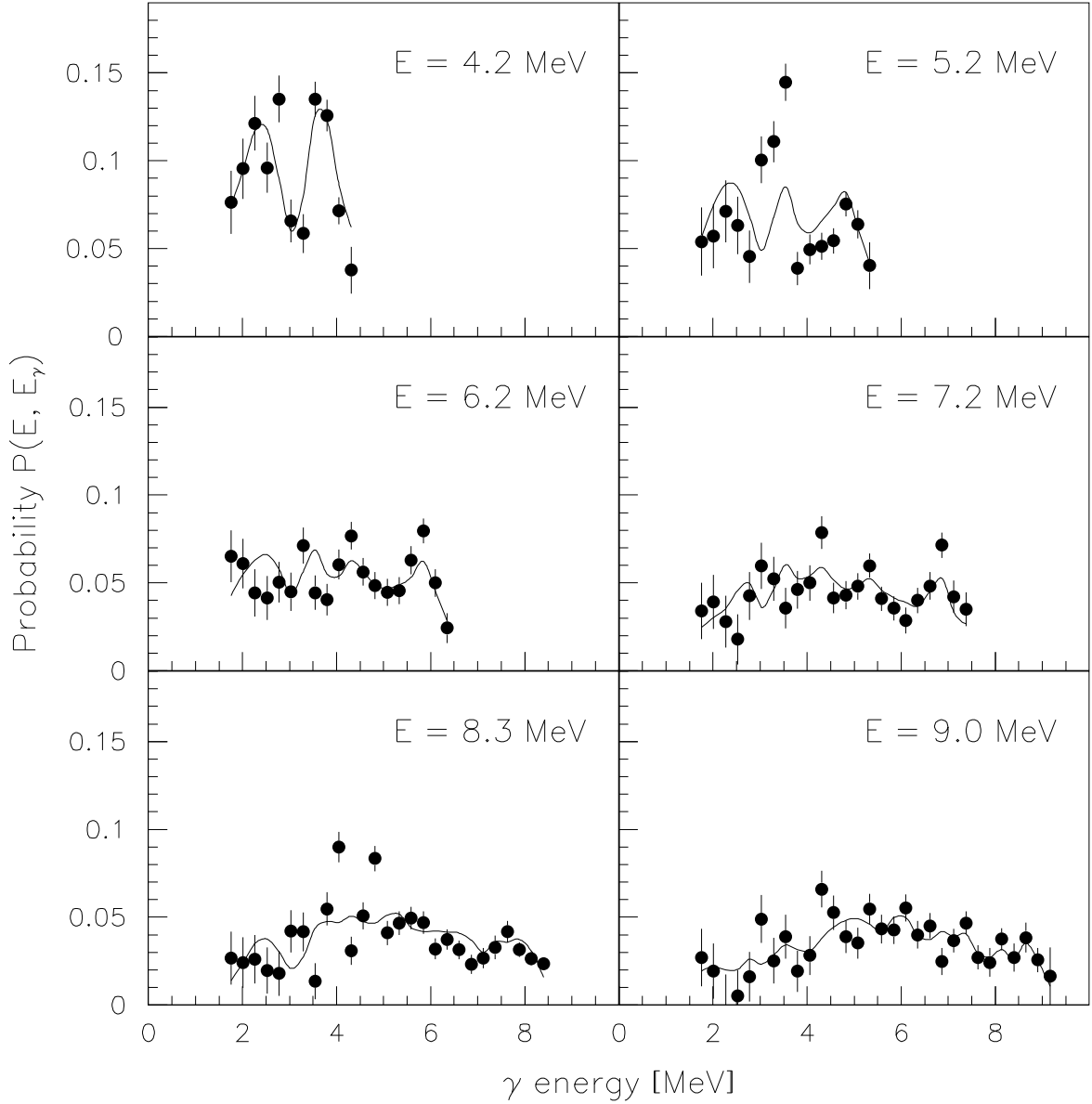}
\caption{Experimental first-generation $\gamma$ spectra (data points with error bars) at six different initial excitation energies (indicated in the figure) compared to the least-$\chi^2$ fit (solid lines) for $^{50}$V. The fit is performed simultaneously on the entire first-generation matrix of which the six displayed spectra are a fraction. The first-generation spectra are normalized to unity for each excitation-energy bin.}
\label{fig:rhosig}
\end{figure}

The $\rho$ and ${\mathcal{T}}$ functions can be determined by an iterative procedure \cite{schi0}, where each data point of these two functions is simultaneously adjusted until a global $\chi^2$ minimum with the experimental $P(E,E_{\gamma})$ matrix is reached. No $a$ $priori$ assumptions about the functional form of either the level density or the $\gamma$-ray transmission coefficient are used. An example to illustrate the quality of the fit is shown in Fig.~\ref{fig:rhosig}, where we compare for the $^{51}$V($^3$He,$\alpha \gamma$)$^{50}$V reaction the experimental first-generation spectra to the least-$\chi^2$ solution for six different initial excitation energies. 

The globalized fitting to the data points determines the functional form of $\rho$ and ${\mathcal{T}}$; however, it has been shown \cite{schi0} that if one solution for the multiplicative functions $\rho$ and ${\mathcal{T}} $ is known, one may construct an infinite number of other functions, which give identical fits to the $P$ matrix by
\begin{eqnarray}
\tilde{\rho}(E-E_\gamma)&=&A\exp[\alpha(E-E_\gamma)]\,\rho(E-E_\gamma),
\label{eq:array1}\\
\tilde{{\mathcal{T}}}(E_\gamma)&=&B\exp(\alpha E_\gamma){\mathcal{T}} (E_\gamma).
\label{eq:array2}
\end{eqnarray}
Thus, the transformation parameters $\alpha$, $A$ and $B$, which correspond to the physical solution, remain to be determined.

\section{Level density and microcanonical entropy}

The parameters $A$ and $\alpha$ can be obtained by normalizing the level density to the number of known discrete levels at low excitation energy~\cite{ENSDF} and to the level density estimated from neutron-resonance spacing data at the neutron binding energy $E=B_n$~\cite{RIPL}. The procedure for extracting the total level density $\rho$ from the resonance energy spacing $D$ is described in Ref.~\cite{schi0}. Since our experimental level-density data points only reach up to an excitation energy of $E\sim B_n-1$ MeV, we extrapolate with the back-shifted Fermi-gas model with a global parametrization~\cite{GC,egidy} 
\begin{equation} 
\rho_{\rm BS}(E)= \eta\frac{\exp(2 \sqrt{aU})}{12 \sqrt{2}a^{1/4}U^{5/4} \sigma_I}, 
\label{eq:bs}
\end{equation}
where a constant attenuation coefficient $\eta$ is introduced to adjust $\rho_{\rm BS}$ to the experimental level density at $B_n$. The intrinsic excitation energy is estimated by $U=E-C_1-E_{\rm pair}$, where $C_1=-6.6A^{-0.32}$ MeV is the back-shift parameter and $A$ is the mass number. The pairing energy $E_{\rm pair}$ is based on pairing gap parameters $\Delta_p$ and $\Delta_n$ evaluated from even-odd mass differences \cite{Wapstra} according to~\cite{BM}. The level-density parameter $a$ and the spin-cutoff parameter $\sigma_I$ are given by $a~=~0.21A^{0.87}~{\rm MeV^{-1}}$ and $\sigma^{2}_{I} = 0.0888TA^{2/3}$, respectively. The nuclear temperature $T$ is described by $T = \sqrt {U/a}$. The parameters used for $^{50,51}$V in Eq.~(\ref{eq:bs}) are listed in Table~\ref{tab:tab1}. 

\begin{figure}
\includegraphics[totalheight=21cm,angle=0,bb=0 80 350 730]{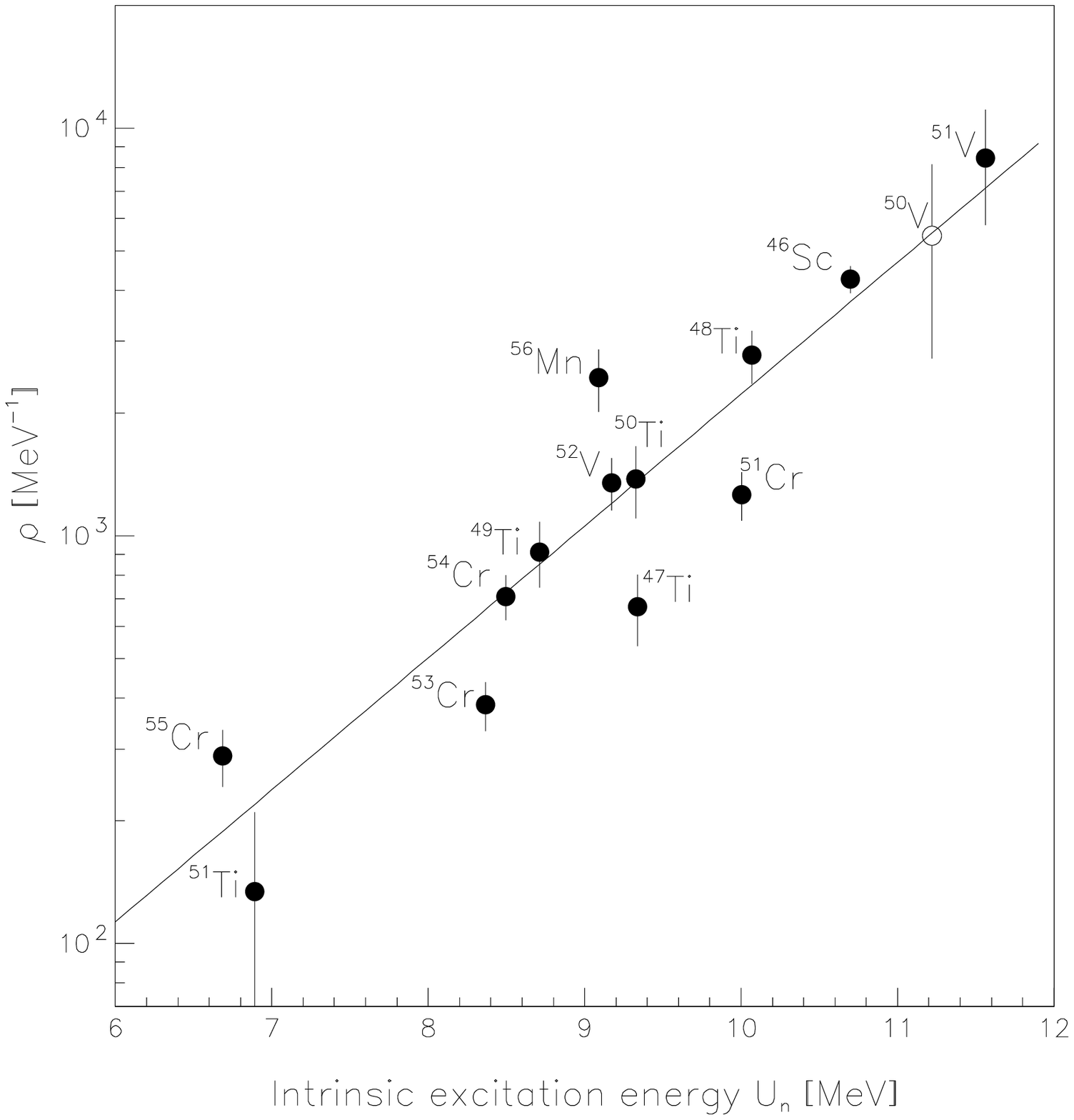}
\caption{Level densities estimated from neutron resonance level spacings at 
$B_n$. The data are plotted as function of intrinsic excitation energy $U_n=B_n-C_1-E_{pair}$. The unknown level density for $^{50}$V (open circle) is estimated from the line determined by a least $\chi^2$ fit to the data points.}
\label{fig:rhosyst}
\end{figure}

\begin{figure}
\includegraphics[totalheight=21cm,angle=0,bb=0 80 350 730]{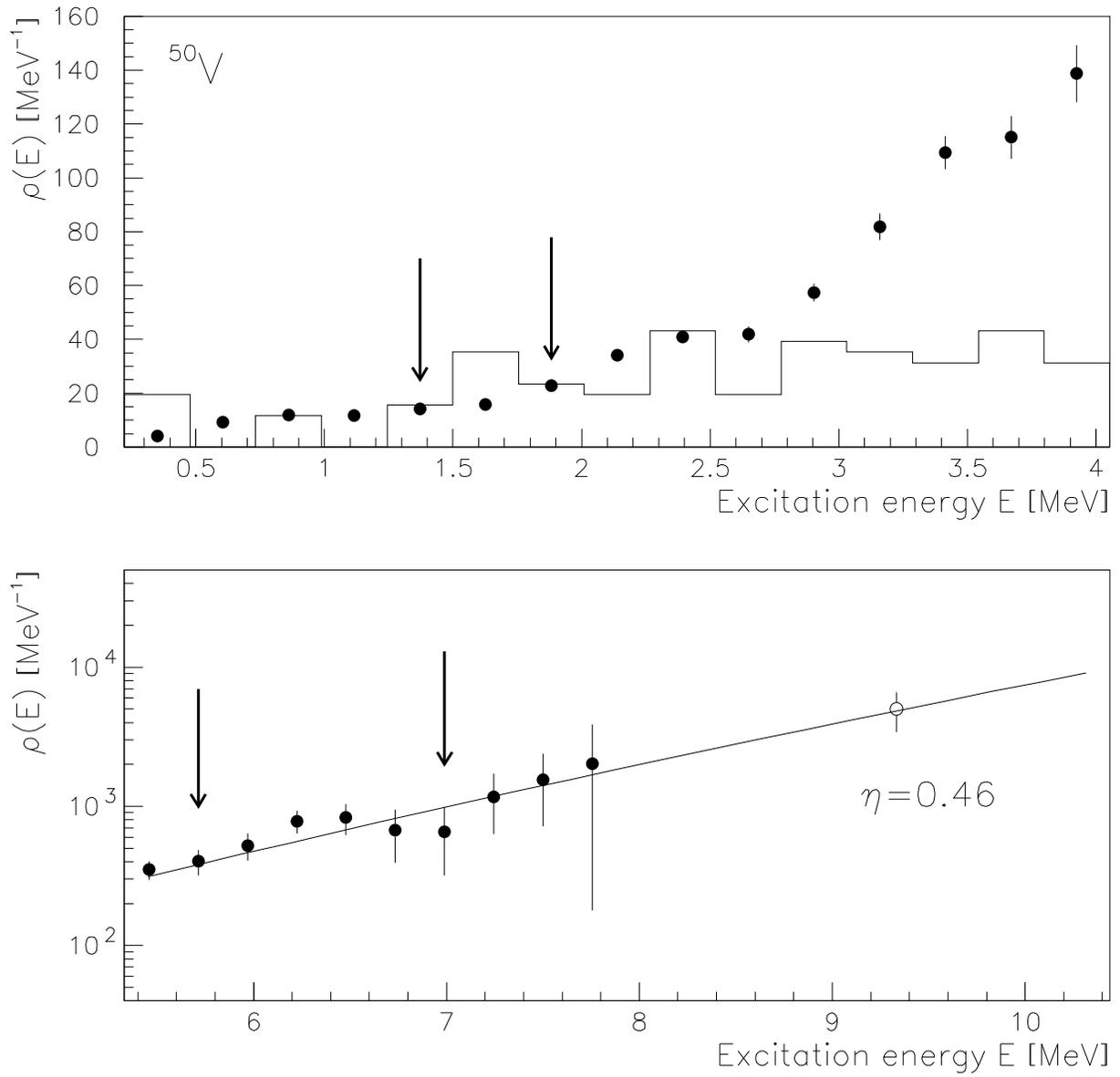}
\caption{Normalization procedure of the experimental level density (data points) of $^{50}$V. The data points between the arrows are normalized to known levels at low excitation energy (histograms) and to the level density at the neutron-separation energy (open circle) using a Fermi-gas level-density extrapolation (solid line).}
\label{fig:counting50v}
\end{figure}

Unfortunately, $^{49}$V is unstable and no information exists on the level density at $E=B_n$ for $^{50}$V. Therefore, we estimate the value from the systematics of other nuclei in the same mass region. In order to bring these data on the same footing, we plot the level densities as a function of intrinsic energy $U$. Due to the strongly scattered data of Fig.~\ref{fig:rhosyst}, the estimate is rather uncertain. We choose a rough estimate of $\rho(B_n)= 5400\pm 2700$ MeV$^{-1}$for $^{50}$V. This value gives an attenuation $\eta =$ 0.46, which is in good agreement with the obtained value of $\eta$ = 0.51 for the $^{51}$V nucleus. Figure \ref{fig:counting50v} demonstrates the level density normalization procedure for the $^{50}$V case, i.e., how the parameters $A$ and $\alpha$ of Eq.~(\ref{eq:array2}) are determined to obtain a level-density function consistent with known experimental data.
 
\begin{figure}
\includegraphics[totalheight=21cm,angle=0,bb=0 0 350 830]{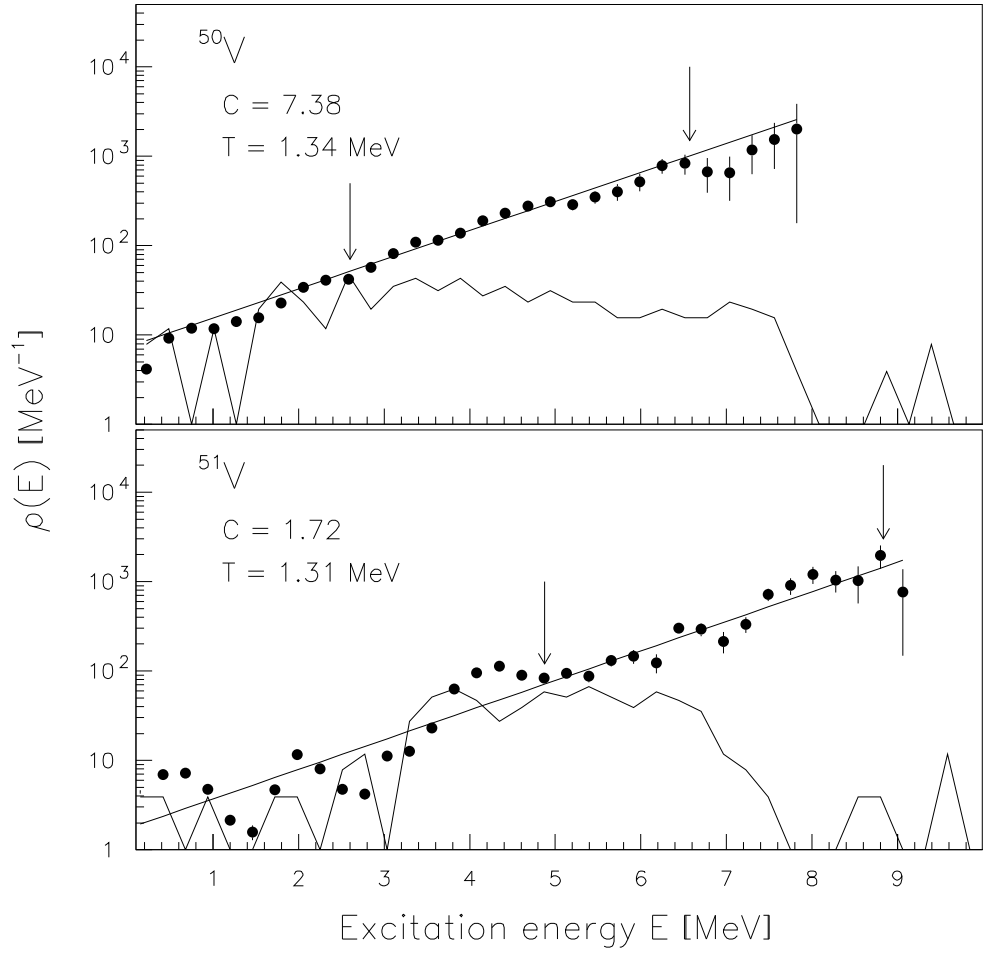}
\caption{Normalized level density of $^{50,51}$V compared to known discrete levels (jagged line) and a constant temperature model (straight line). The fits are performed in the region between the arrows.}
\label{fig:rholev}
\end{figure}

The experimentally extracted and normalized level densities of $^{50}$V and $^{51}$V are shown in Fig.~\ref{fig:rholev} for excitation energies up to $\sim$ 8 and 9 MeV, respectively. The level density of $^{50}$V is relatively high and has a rather smooth behaviour due to the effect of the unpaired proton and neutron, while the level density of $^{51}$V displays distinct structures for excitation energies up to $\sim$ 4.5 MeV. This effect is probably caused by the closed $f_{7/2}$ neutron shell in this nucleus. 

The level densities of $^{50,51}$V obtained with the Oslo method are compared to the number of levels from spectroscopic experiments~\cite{ToI}. The $^{51}$V nucleus has relatively few levels per energy bin because of its closed neutron shell, so using spectroscopic methods to count the levels seems to be reliable up to $\sim$ 4 MeV excitation energy in this case. For higher excitations the spectroscopic data are significantly lower compared to the level density obtained with the Oslo method. This means that many levels are not accounted for in this excitation region by using standard methods. The same can be concluded for $^{50}$V, and in this case the spectroscopic level density drops off already at an excitation energy of about 2.5 MeV. 

The level densities of $^{50,51}$V are also compared to the constant-temperature formula
\begin{equation}
\rho_{\rm fit}= C \exp (E/T),
\label{eq:ct}
\end{equation}
which is drawn as a solid line in Fig.~\ref{fig:rholev}. Here the parameters $C$ and $T$ are the level density at about zero excitation energy and the average temperature, respectively; both are estimated from the fit of the exponential to the region of the experimental level density indicated by arrows. From this model a constant temperature of about 1.3 MeV is found for both nuclei. 

The level density of a system can give detailed insight into its thermal properties. The multiplicity of states $\Omega_{s}(E)$, which is the number of physically obtainable realizations available at a given energy, is directly proportional to the level density and a spin-dependent factor $(2\langle J(E) \rangle +1)$, thus
\begin{equation}
\Omega_{s}(E)\propto \rho (E) (2\langle J(E)\rangle +1),
\end{equation}
where $\langle J(E)\rangle$ is the average spin at excitation energy $E$. Unfortunately, the experimentally measured level density in this work does not correspond to the true multiplicity of states, since the $(2J+1)$ degeneracy of magnetic substates is not included. If the average spin of levels $\langle J\rangle$ at any excitation energy were known, this problem could be solved by multiplying an energy-dependent factor $(2\langle J(E) \rangle+1)$ to the experimental level density. However, little experimental data exist on the spin distribution. Therefore, we choose in this work to use a multiplicity $\Omega_{l}(E)$ based on the experimental level density alone:
\begin{equation}
\Omega_{l}(E) \propto  \rho (E).
\end{equation}

The entropy $S(E)$ is a measure of the degree of disorder of a system at a specific energy. The microcanonical ensemble, where the system is completely isolated from any exchange with its surroundings, is often considered as the appropriate one for the atomic nucleus since the strong force has such a short range, and the nucleus normally does not share its excitation energy with the external environment. 

According to our definition of the multiplicity of levels $\Omega_{l}(E)$ obtained from the experimental level density, we define a "pseudo" entropy 
\begin{equation}
S(E)=k_{\rm B}\ln \Omega_{l}(E),
\end{equation}
which is utilized in the following discussion. For convenience Boltzmann's constant $k_{B}$ can be set to unity.

In order to normalize the entropy the multiplicity is written as $\Omega_{l}(E)= \rho(E)/\rho_0$. The normalization denominator $\rho_0$ is to be adjusted such that the entropy approaches a constant value when the temperature approaches zero in order to fullfill the third law of thermodynamics: $S(T\rightarrow 0) = S_{0}$. In the case of even-even nuclei the ground state represents a completely ordered system with only one possible configuration. This means that the entropy in the ground state is $S = \ln 1 = 0$, and the normalization factor $1/\rho_0$ is chosen such that this is the case. Since the vanadium nuclei have an odd number of protons, a $\rho_0$ which is typical for even-even nuclei in this mass region is used for both the $^{50}$V and the $^{51}$V nucleus. The normalization factor $\rho_0$ used is 0.7 MeV$^{-1}$, found from averaging data on $^{50}$Ti and $^{52}$Cr. 

\begin{figure}
\includegraphics[totalheight=21cm,angle=0,bb=0 0 350 830]{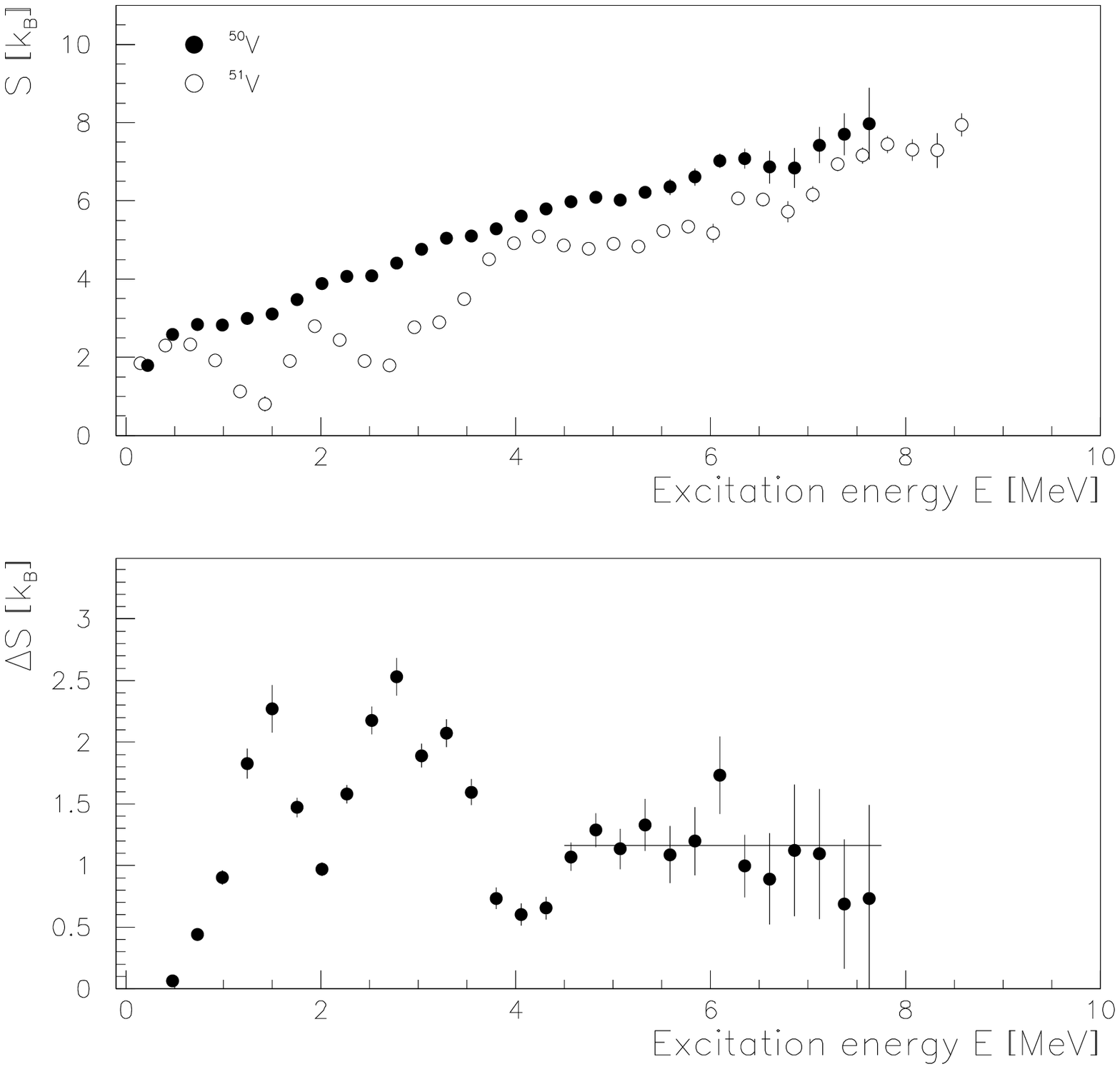}
\caption{Entropies of $^{50,51}$V (upper panel), entropy difference between the two vanadium isotopes (lower panel).}
\label{fig:vds}
\end{figure}

The entropies of $^{50,51}$V extracted from the experimental level density are shown in the upper panel of Fig.~\ref{fig:vds}. Naturally, they show the same features as the level density plot, with the odd-odd $^{50}$V displaying higher entropy than the odd-even $^{51}$V. Since the neutrons are almost ($^{50}$V) or totally ($^{51}$V) blocked at low excitation energy, the multiplicity and thus the entropy is made primarily by the protons in this region. 

At 4 MeV of excitation energy a relatively large increase of the entropy is found in the case of $^{51}$V. This is probably because the excitation energy is large enough to excite a nucleon across the $N,Z = 28$ shell gap to other orbitals.  

In the excitation region above $\sim$ 4.5 MeV the entropies show similar behaviour, which is also expressed by the entropy difference $\Delta S$ displayed in the lower panel of Fig.~\ref{fig:vds}. We assume here that the two systems have an approximately statistical behaviour, and that the neutron hole in $^{50}$V acts as a spectator to the $^{51}$V core. The entropy of the hole can be estimated from the entropy difference $\Delta S = S$($^{50}$V)$-S$($^{51}$V). From the lower panel of Fig.~\ref{fig:vds} we find $\Delta S \sim 1.2 k_B$ for $E > 4.5$ MeV. This is slightly less than the quasi-particle entropy found in rare-earth nuclei, which is estimated to be $\Delta S \sim 1.7 k_B$~\cite{gutt3}. This is not unexpected since the single-particle levels are more closely spaced for these nuclei; they have therefore more entropy. 

The naive configurations of $^{50,51}$V at low excitations are $\pi f_{7/2}^{3} \nu f_{7/2}^{7}$ and $\pi f_{7/2}^{3} \nu f_{7/2}^{8}$, respectively. Thus, by counting the possible configurations within the framework of the BCS model~\cite{BCS} in the nearly degenerate $f_{7/2}$ shell, one can estimate the multiplicity of levels and thus the entropy when no Cooper pairs are broken in the nucleus, one pair is broken and so on. We assume a small deformation that gives four energy levels with Nilsson quantum number $\Omega = 1/2,\, 3/2,\, 5/2,\, 7/2$. Furthermore, we neglect the proton-neutron coupling and hence assume that the protons and the neutrons can be considered as two separate systems; the total entropy based on the number of energy levels can then be written as $S = S_p + S_n$. This gives $S = 2.8 k_B$ for the nucleus $^{50}$V, and $S = 1.4 k_B$ for $^{51}$V when two protons are coupled in a Cooper pair. These values are in fair agreement with the data of Fig.~\ref{fig:vds} at an excitation energy below $\sim 2$ MeV. It is gratifying that these crude estimates give an entropy of the neutron hole in $^{50}$V of $\Delta S = 1.4 k_B$, in good agreement with the experimental value for the entropy difference of $1.2 k_B$ found from Fig.~\ref{fig:vds}.

With the three $f_{7/2}$ protons unpaired we obtain a total entropy of $S = 3.5$ and $2.1 k_B$ for $^{50,51}$V, respectively. This means that the process of just breaking a proton pair within the same shell does not contribute much to the total entropy, but when a nucleon has enough energy to cross the shell gap a significant increase of the entropy is expected. As already mentioned, at excitation energies above $\sim$ 4 MeV, it is very likely that configurations from other shells will participate in building the total entropy.

\section{Radiative strength functions}

The $\gamma$-ray transmission coefficient ${\mathcal{T}}$ in Eq.~(\ref{eq:axel}) is expressed as a sum of all the RSFs $f_{XL}$ of electromagnetic character $X$ and multipolarity $L$:
\begin{equation}
{\mathcal{T}}(E_{\gamma}) = 2\pi \sum_{XL} E_{\gamma}^{2L+1} f_{XL}(E_{\gamma}).
\end{equation}
The slope of the experimental $\gamma$-ray transmission coefficient ${\mathcal{T}}$ has been determined through the normalization of the level densities, as described in Sect.~III. The remaining constant $B$ in Eq.~(\ref{eq:array2}) is determined using information from neutron resonance decay, which gives the absolute normalization of ${\mathcal{T}}$. For this purpose we utilize experimental data~\cite{RIPL} on the average total radiative width $\langle\Gamma_{\gamma} \rangle$ at $E=B_{n}$. 

We assume here that the $\gamma$-decay taking place in the quasi-continuum is dominated by $E1$ and $M1$ transitions and that the number of positive and negative parity states is equal. For initial spin $I$ and parity $\pi$ at $E=B_n$, the expression of the width~\cite{kopecky} reduces to
\begin{eqnarray}
\langle\Gamma_\gamma\rangle=\frac{1}{4 \pi \rho(B_n, I, \pi)} \sum_{I_f}&&\int_0^{B_n}{\mathrm{d}}E_{\gamma} B{\mathcal{T}} (E_{\gamma})
\nonumber\\
&&\rho(B_n-E_{\gamma}, I_f),
\label{eq:norm}
\end{eqnarray}
where $D_i = 1/\rho(B_n,I,\pi)$ is the average spacing of $s$-wave neutron resonances. The summation and integration run over all final levels with spin $I_f$, which are accessible by dipole ($L=1$) $\gamma$ radiation with energy $E_{\gamma}$. From this expression the normalization constant $B$ can be determined as described in Ref.~\cite{voin1}. However, some considerations have to be made before normalizing according to Eq.~(\ref{eq:norm}). 

\begin{figure}
\includegraphics[totalheight=21cm,angle=0,bb=0 80 350 730]{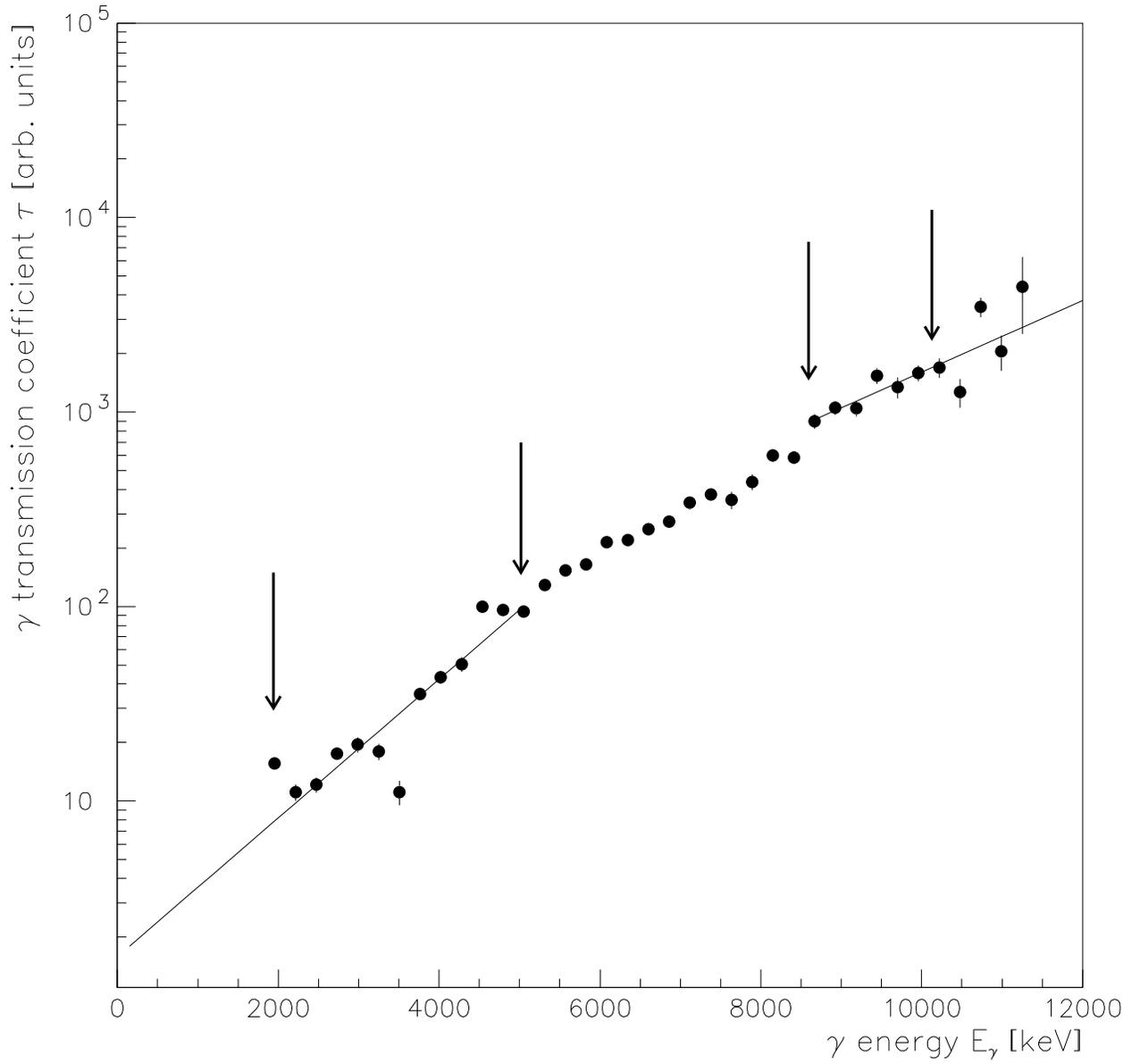}
\caption{Unnormalized $\gamma$-ray transmission coefficient for $^{51}$V. The lines are extrapolations needed to calculate the normalization integral of Eq.~(\ref{eq:norm}). The arrows indicate the lower and upper fitting regions for the extrapolations.}
\label{fig:sigext}
\end{figure}

Methodical difficulties in the primary $\gamma$-ray extraction prevent determination of the function ${\mathcal{T}}(E_{\gamma})$ in the interval $E_\gamma<1$~MeV. In addition, the data at the highest $\gamma$-energies, above $E_{\gamma} \sim B_n-1$~MeV, suffer from poor statistics. We therefore extrapolate ${\mathcal{T}}$ with an exponential form, as demonstrated for $^{51}$V in Fig.~\ref{fig:sigext}. The contribution of the extrapolation to the total radiative width given by Eq.~(\ref{eq:norm}) does not exceed $15$\%, thus the errors due to a possibly poor extrapolation are expected to be of minor importance~\cite{voin1}. 

Again, difficulties occur when normalizing the $\gamma$-ray transmission coefficient in the case of $^{50}$V due to the lack of neutron resonance data. Since the average total radiative width $\langle\Gamma_{\gamma} \rangle$ at $E=B_{n}$ does not seem to show any clear systematics for nuclei in this mass region, we choose the same absolute value of the GEDR tail for $^{50}$V as the one found for $^{51}$V from photoabsorption experiments. The argument for this choice is that the GEDR should be similar for equal number of protons provided that the two nuclei have the same shapes. 

Since it is assumed that the radiative strength is dominated by dipole transitions, the RSF can be calculated from the normalized transmission coefficient by
\begin{equation}
f(E_{\gamma}) =\frac{1}{2\pi}\frac{ {\mathcal{T}} (E_{\gamma}) }{ E_{\gamma}^{3} }.
\end{equation}
We would now like to decompose the RSF into its components from different multipolarities to investigate how the $E1$ and $M1$ radiation contribute to the total strength. 

The Kadmenski{\u{\i}}, Markushev and Furman (KMF) model~\cite{kad} is employed for the $E1$ strength. In this model, the Lorentzian GEDR is modified in order to reproduce the nonzero limit of the GEDR for $E_{\gamma} \rightarrow 0$ by means of a temperature-dependent width of the GEDR. The $E1$ strength in the KMF model is given by
\begin{equation} 
f_{E1}(E_\gamma)=\frac{1}{3\pi^2\hbar^2c^2} \frac{0.7\sigma_{E1}\Gamma_{E1}^2(E_\gamma^2+4\pi^2T^2)} {E_{E1}(E_\gamma^2-E_{E1}^2)^2},
\label{eq:E1}
\end{equation}
where $\sigma_{E1}$ is the cross section, $\Gamma_{E1}$ is the width, and $E_{E1}$ is the centroid of the GEDR determined from photoabsorbtion experiments. 

We adopt the KMF model with the temperature $T$ taken as a constant to be consistent with our assumption that the RSF is independent of excitation energy. The possible systematic uncertainty caused by this assumption is estimated to have a maximum effect of 20\% on the RSF~\cite{gutt7}. The values used for $T$ are the ones extracted from the constant-temperature model in Eq.~(\ref{eq:ct}). 

The GEDR is split into two parts for deformed nuclei. Data of $^{51}$V from photoabsorption experiments show that the GEDR is best fitted with two Lorentzians, indicating a splitting of the resonance and a non-zero ground-state deformation of this nucleus. Indeed, $B$($E2$) values~\cite{RIPL} suggest a deformation of $\beta \sim$ 0.1 for $^{50,51}$V. Therefore, a sum of two modified Lorentzians each described by Eq.~(\ref{eq:E1}) is used (see Table~\ref{tab:tab2}). 

For $f_{M1}$, which is supposed to be governed by the spin-flip $M1$ resonance \cite{voin1}, the Lorentzian giant magnetic dipole resonance (GMDR)
\begin{equation}
f_{M1}(E_\gamma)=\frac{1}{3\pi^2\hbar^2c^2} \frac{\sigma_{M1}E_\gamma\Gamma_{M1}^2} {(E_\gamma^2-E_{M1}^2)^2+E_\gamma^2\Gamma_{M1}^2}
\label{eq:M1}
\end{equation}
is adopted. 

\begin{figure}
\includegraphics[totalheight=21cm,angle=0,bb=0 10 350 830]{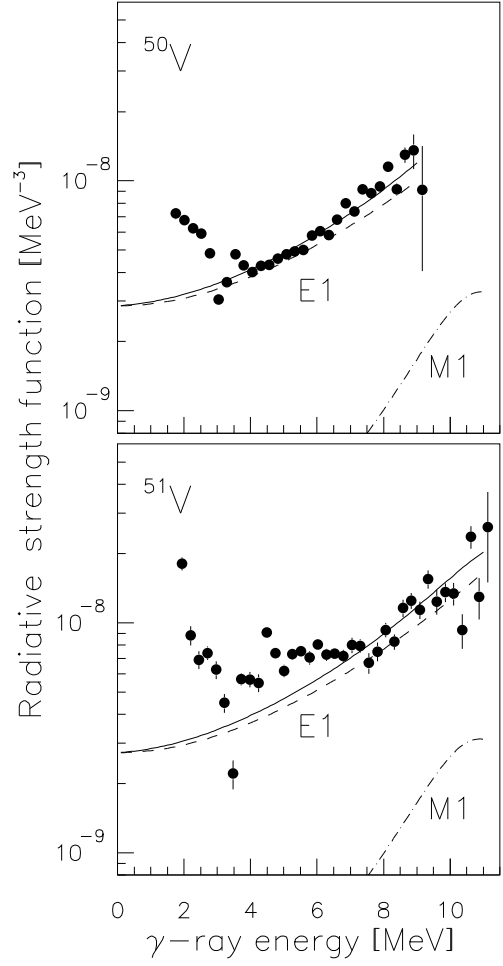}
\caption{Normalized RSFs of $^{50,51}$V. The dashed and dash-dotted line show the extrapolated tails of the giant electric and giant magnetic dipole resonance, respectively. The solid line is the summed strength for the giant dipole resonances.}
\label{fig:strengthboth}
\end{figure}

The GEDR and GMDR parameters are taken from the systematics of Ref.~\cite{RIPL} and are listed in Table~\ref{tab:tab2}. Thus, we fit the total RSF given by
\begin{equation}
f=\kappa (f_{E1,1} + f_{E1,2} + f_{M1})
\end{equation}
to the experimental data using the normalization constant $\kappa$ as a free parameter. The value of $\kappa$ generally deviates from unity due to theoretical uncertainties in the KMF model and the evaluation of the absolute normalization in Eq.~(\ref{eq:norm}). The resulting RSFs extracted from the two reactions are displayed in Fig.~\ref{fig:strengthboth}, where the data have been normalized with parameters from Tables~\ref{tab:tab1} and~\ref{tab:tab2}. 

The $\gamma$-decay probability is governed by the number and the character of available final states and by the RSF. A rough inspection of the experimental data of Fig.~\ref{fig:strengthboth} indicates that the RSFs are increasing functions of $\gamma$-energy, generally following the tails of the GEDR and GMDR resonances in this region. 

At low $\gamma$ energies ($E_{\gamma} \lesssim 3$ MeV), an enhancement of a factor of $\sim$ 5 over the KMF estimate of the strength appears in the RSFs. This increase has also been seen in some Fe~\cite{voinov} and Mo~\cite{gutt7} isotopes, where it has been shown to be present in the whole excitation-energy region. In the case of the $^{57}$Fe RSF, the feature has been confirmed by an (n,$2\gamma$) experiment~\cite{voinov}. However, it has not appeared in the RSFs of the rare-earth nuclei investigated earlier by the Oslo group. The physical origin of the enhancement has not, at present, any satisfying explanation, as none of the known theoretical models can account for this behaviour.

So far, we have not been able to detect any technical problems with the Oslo method. The unfolding procedure with the NaI response functions gives reliable results, as demonstrated in Fig.~\ref{fig:unfolding}. Also, Fig.~\ref{fig:gamma} indicates that the low-energy $\gamma$ intensity is subtracted correctly; if not, one would find less intensity in the higher-generations spectrum at these $\gamma$ energies. Figure~\ref{fig:rhosig} shows the final test, where the result from the least-$\chi^2$ fit nicely reproduces the experimental data. In addition, investigations in $^{27,28}$Si~\cite{Si} showed that our method produced $\gamma$-transition coefficients in excellent agreement with average decay widths of known, discrete transitions. Hence, we do not believe that the enhancement is caused by some technical or methodical problems. Still, independent confirmation of the increasing RSF from, e.g., (n,$2\gamma$) experiments on the V and Mo isotopes, is highly desirable.

\section{Summary and conclusions}

The Oslo method has been applied to extract level densities and RSFs of the vanadium isotopes $^{50,51}$V. From the measured level densities the microcanonical entropies have been derived. The entropy carried by the neutron hole in $^{50}$V is estimated to be $\sim$ 1.2 $k_B$, which is less than the quasi-particle entropy of $\sim$ 1.7 $k_B$ found in rare-earth nuclei. 

The experimental RSFs are generally increasing functions of $\gamma$ energy. The main contribution to the RSFs is the GEDR; also the GMDR is present. At low $\gamma$ energies an increase of the strength functions is apparent. A similar enhancement has also been seen in the iron and molybdenum isotopes. There is still no explanation for the physics behind this very interesting behaviour.

\acknowledgements
Financial support from the Norwegian Research Council (NFR) is gratefully acknowledged. A.V. acknowledges support from a NATO Science Fellowship under project number 150027/432 given by the Norwegian Research Council (NFR). A.V. also acknowledges support from Stewardship Science Academic Alliances, grant number DE-FG03-03-NA0074.

\end{multicols}

\newpage

\begin{table}
\caption{Parameters used for the back-shifted Fermi gas level density.} 
\begin{tabular}{l|ccc|ccc|c}
Nucleus    & $E_{\rm pair}$ & $a$         & $C_1$ &  $B_n$ & $D     $ & $\rho (B_n)$ &$\eta$ \\ 
           &     (MeV)      & (MeV$^{-1}$)& (MeV) & (MeV)  &   (keV)   & ($10^3$MeV$^{-1}$) & \\
\hline
&&&&&&\\

$^{50}$V &     0      & 6.31       & -1.89 & 9.33 &-       &5.4(16)$^a$ & 0.46     \\
$^{51}$V &     1.36      & 6.42       & -1.88 & 11.05  &2.3(6)&8.4(26) & 0.51     \\
\end{tabular}
$^a$Estimated from systematics.
\label{tab:tab1}
\end{table}

\begin{table}
\caption{Parameters used for the radiative strength functions.} 
\begin{tabular}{l|ccc|ccc|ccc|ccc}
Nucleus &$E_{{\mathrm{E}}1,1}$&$\sigma_{{\mathrm{E}}1,1}$&$\Gamma_{{\mathrm{E}}1,1}$&$E_{{\mathrm{E}}1,2}$&$\sigma_{{\mathrm{E}}1,2}$&$\Gamma_{{\mathrm{E}}1,2}$&$E_{{\mathrm{M}}1}$&$\sigma_{{\mathrm{M}}1}$&$\Gamma_{{\mathrm{M}}1}$&$\langle \Gamma_\gamma \rangle$&$T$&$\kappa$\\
        &(MeV)     &(mb)           &(MeV)          &(MeV)     &(mb)           &(MeV)          &(MeV)   &(mb)         &(MeV)        &(meV) &(MeV) & \\ \hline
&&&&&&&&&&\\
$^{50}$V& 17.93& 53.3& 3.62& 20.95& 40.7& 7.15& 11.1& 0.532& 4.0& - & 1.34& 0.75\\
$^{51}$V& 17.93& 53.3& 3.62& 20.95& 40.7& 7.15& 11.1& 0.563& 4.0& 600(80)& 1.31& 0.74\\
\end{tabular}
\label{tab:tab2}
\end{table}

\clearpage

\end{document}